# Strange Quark Physics on the Lattice

*UKQCD Collaboration*
presented by David Richards[a]

[a]Department of Physics & Astronomy, University of Edinburgh, Edinburgh EH9 3JZ, Scotland

We present results for hadrons containing a strange quark in quenched lattice QCD. We calculate masses and decay constants using 60 gauge configurations with an $O(a)$-improved fermion action at $\beta = 6.2$. Using the $\rho$ mass to set the scale, we find hadron masses within two to three standard deviations of experiment. Direct comparison with experiment for decay constants is obscured by uncertainty in current renormalisations.

Finally, we present preliminary results on the semi-leptonic decay $D \to K$. We find $f_K^+(0)/Z_V = 0.75 \, ^{+5}_{-4}$ and $f_K^0(0)/Z_V = 0.74 \, ^{+4}_{-4}$.

## 1. Introduction

Current quenched lattice simulations employ a box with a linear size of around 2 fm, precluding the study of the $u$ and $d$ quarks directly. However, such lattices do allow the direct simulation of the strange quark, thereby revealing a significantly wider range of physical quantities with which to explore lattice QCD. In this talk, I will discuss the extent to which we are able to compute the masses and leptonic decay rates of hadrons containing the $s$ quark - "benchmarking" our lattice calculations. I will then present some preliminary results on semi-leptonic decays into hadrons containing the $s$ quark: the $D \to K$ form factors.

It is generally assumed that the masses, $m_P$ and $m_V$, of the pseudoscalar and vector mesons respectively obey the relations

$$m_P^2(m_1, m_2) = b_P(m_1 + m_2) \qquad (1)$$
$$m_V(m_1, m_2) = a_V + b_V(m_1 + m_2) \qquad (2)$$

where $m_1$ and $m_2$ are the quark masses. The quark mass, $m$, is related to the hopping parameter, $\kappa$, and its critical value, $\kappa_{\rm crit}$, by

$$m = \frac{1}{2}\left(\frac{1}{\kappa} - \frac{1}{\kappa_{\rm crit}}\right). \qquad (3)$$

We will test the assumptions (1) and (2) by fitting the lattice data for $m_P^2, m_V, f_P$, etc. to a more general function of the quark masses, $m_1$ and $m_2$:

$$a_1 + \frac{a_2}{2}(m_1 + m_2) + \frac{a_3}{2}|m_2 - m_1|. \qquad (4)$$

By a mixture of extrapolation and interpolation using this function, we will compute the masses and decay constants of the $K, K^*$ and $\Phi$, in addition to those of the usual light hadrons.

Having determined $\kappa_s$, we will then proceed to study the semi-leptonic decays of the $D$ meson into the $K$, and our expectation for the behaviour of the corresponding form factors with quark mass and momentum.

## 2. Simulation Details

We have analysed 60 gauge field configurations at $\beta = 6.2$ on $24^3 \times 48$ lattices. $O(a)$-improved "clover" propagators were computed at three "light" values of $\kappa$, 0.14144, 0.14226, 0.14262, and at $\kappa_c = 0.129$, corresponding very closely to the mass of the charm quark. The light-quark propagators were computed using local sources and sinks; the "charm" propagator was computed using a gauge-invariant "smeared" source to both local and smeared sinks. Full details of the calculations are contained in refs. [1,2]. Correlations between the data at different timeslices are maintained in the fits, with the errors extracted using a bootstrap procedure.

## 3. Masses and leptonic widths

We compute meson correlators for both degenerate and nondegenerate quark masses. Fits of the meson masses and the decay constants $f_P$

Table 1
Parameters of fit to Equation (4) for masses, decay constants and ratios.

|  | $a_1$ | $a_2$ | $a_3$ |
|---|---|---|---|
| $m_P^2$ | 0.0 | $2.12\,^{+4}_{-3}$ | $-0.01\,^{+1}_{-1}$ |
| $m_V$ | $0.29\,^{+2}_{-1}$ | $2.3\,^{+4}_{-3}$ | $0.1\,^{+1}_{-1}$ |
| $\frac{f_P}{Z_A}$ | $0.041\,^{+1}_{-2}$ | $0.53\,^{+3}_{-2}$ | $-0.01\,^{+1}_{-1}$ |
| $\frac{1}{f_V Z_V}$ | $0.38\,^{+1}_{-2}$ | $-1.4\,^{+4}_{-2}$ | $0.0\,^{+1}_{-1}$ |
| $\frac{f_P}{m_V Z_A}$ | $0.142\,^{+6}_{-9}$ | $0.5\,^{+2}_{-2}$ | $-0.10\,^{+6}_{-7}$ |

and $1/f_V$ to Equation (4) are presented in Table 1. The values obtained for $a_3$ are consistent with zero in all cases, and the $\chi^2/\text{dof}$ are satisfactory. As there is no theoretical justification for $a_3 \neq 0$, we will henceforth present results with $a_3$ constrained to be zero. Using this fit, we obtain $\kappa_{\text{crit}} = 0.14315\,^{+2}_{-2}$, and a value for the inverse lattice spacing, $a^{-1}$, of $2.7\,^{+1}_{-1}$ GeV, consistent with the value of 2.73(5) GeV from the string tension.

We compute the baryon correlators only with degenerate quark masses. The quality of the chiral extrapolation for the nucleon is rather poor, with a $\chi^2/\text{dof}$ of 5.1, and we obtain $a^{-1} = 3.0\,^{+2}_{-3}$ GeV, somewhat larger than the values above. A linear fit to $m_N^2$ provides a smaller $\chi^2/\text{dof}$ of 2.7, and a still larger value of $a^{-1}$. The linear fit to $m_\Delta$ is perfectly acceptable, though with large errors.

We determine $\kappa_s$, corresponding to the $s$ quark mass, from a fit of $m_P^2(\kappa_1, \kappa_2)$ to the physical value of $m_K^2$ using the scale $a^{-1}(m_\rho)$, yielding $\kappa_s = 0.1429\,^{+1}_{-1}$. We can use this value to determine the renormalized strange quark mass in the $\overline{\text{MS}}$ scheme, and obtain $m_s^{\overline{\text{MS}}}(2\,\text{GeV}) = 109\,^{+11}_{-11}$.

We present the results for meson masses and decay constants, extrapolated to physical $\kappa$ values, in Table 2. For the decay constants, we use the "boosted" coupling [3]. The masses are generally within $2-3\sigma$ of their physical values. However, the pseudoscalar decay constants are $5-7\sigma$ below their physical values. This is illustrated in Figure 1; the discrepancy may be due to the effects of quenching, to the use of the perturbative value of $Z_A$, or some combination.

Table 2
Masses and decay constants, using $a^{-1}(m_\rho)$.

|  | lattice |  | experiment |  |
|---|---|---|---|---|
| $m_{\eta_s}$ | $670\,^{+10}_{-10}$ | MeV | "686 | MeV" |
| $m_{K^*}$ | $868\,^{+9}_{-8}$ | MeV | 892 | MeV |
| $m_\phi$ | $970\,^{+20}_{-10}$ | MeV | 1020 | MeV |
| $f_\pi$ | $102\,^{+6}_{-7}$ | MeV | 132 | MeV |
| $f_K$ | $123\,^{+5}_{-6}$ | MeV | 160 | MeV |
| $1/f_\rho$ | $0.316\,^{+7}_{-13}$ |  | 0.28 |  |
| $1/f_{K^*}$ | $0.298\,^{+5}_{-9}$ |  |  |  |
| $1/f_\phi$ | $0.280\,^{+3}_{-6}$ |  | 0.23 |  |
| $m_N$ | $820\,^{+90}_{-60}$ | MeV | 938 | MeV |
| $m_\Delta$ | $1300\,^{+100}_{-100}$ | MeV | 1232 | MeV |

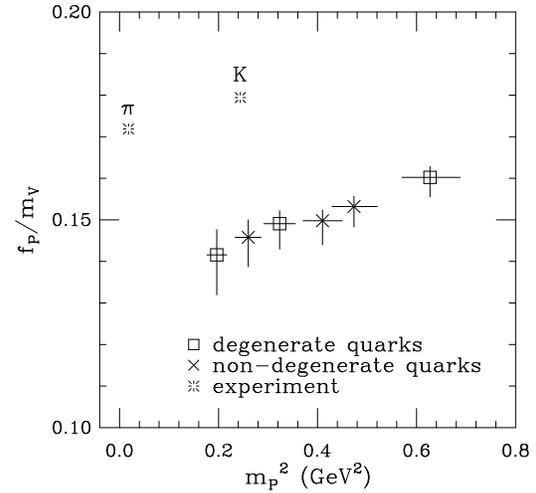

Figure 1. $f_P/m_V$ vs. $m_P^2$, together with the physical values. The perturbative value of $Z_A$ using the boosted coupling is used.

## 4. $D \to K$ form factors

The decay is described in terms of two form factors, $f_K^+(q^2)$ and $f_K^0(q^2)$:

$$<K|V_\mu|D> = (p_D + p_K - \frac{m_D^2 - m_K^2}{q^2}q)_\mu f_K^+(q^2)$$

$$+ \frac{m_D^2 - m_K^2}{q^2} q_\mu f_K^0(q^2). \qquad (5)$$

The $D$ meson is placed at $t = 24$, the mid-point of the lattice, and has momentum $|\vec{p}_D| = 0, \frac{\pi}{12}a^{-1}$. Momentum $\vec{q}$ is inserted at the operator, with $|\vec{q}| \leq \frac{\pi}{12}2a^{-1}$. We use the local, improved vector

Table 3
Intercept and masses from fit to Equation (7).

| $f_K^+(0)/Z_V$ | $m_1$ | $f_K^0(0)/Z_V$ | $m_0$ |
|---|---|---|---|
| $0.74\,^{+5}_{-4}$ | $0.7\,^{+1}_{-1}$ | $0.74\,^{+4}_{-4}$ | $1.0\,^{+1}_{-1}$ |

current.

We perform a fit to the three-point correlators, constraining the energies and the matrix elements $<0|\bar{\psi}_1\gamma_5\psi_2|P>$ from fits to appropriate two-point functions. For the $K$ meson, for which we use a local interpolating operator, the energies at finite momentum are computed using the mass together with the continuum dispersion relation, and we employ the wave function determined from the zero-momentum two-point function. For the $D$ meson, where we use a smeared interpolating operator, we cannot do this.

Denoting the active- and spectator-quark hopping parameters by $\kappa_1$ and $\kappa_2$ respectively, we extrapolate the form factors at fixed $\vec{p}_D$ and $\vec{p}_K$, $F(\kappa_1,\kappa_2)$, to the physical $\kappa$'s as follows. We extrapolate $\kappa_2$ to $\kappa_{\rm crit}$ according to:

$$F(\kappa_1,\kappa_2) = F(\kappa_1,\kappa_{\rm crit}) + A(\frac{1}{\kappa_2} - \frac{1}{\kappa_{\rm crit}}). \qquad (6)$$

We then interpolate between $\kappa_1 = 0.14144$ and $0.14226$ to the value of $\kappa_s$ determined above. $q^2$ is computed using the extrapolated masses. Note that, in contrast to some analyses [4,5], we do not assume flavour symmetry between the active and spectator quarks. Both the extrapolation and interpolation use the full covariance matrix.

We perform a correlated, two-parameter fit of the form factors to a meson-dominance model,

$$f_K^{+,0}(q^2) = f_K^{+,0}(0)/(1 - \frac{q^2}{m_l^2}). \qquad (7)$$

Our expectation is that $m_l$ should be the $t$-channel mass. Only those points for which $|\vec{p}_K| \le \frac{\pi}{12}a^{-1}$ are included in the fit. The masses and intercepts are shown in Table 3, and the quality of the extrapolation for $f_K^{+,0}(q^2)$ is illustrated in Figure 2. The values obtained for the intercept are in broad agreement with those obtained by other groups [4–8].

The principal systematic uncertainty is in the matching factor, $Z_V$ [5,8]. Though we are un-able to measure the form factors using the conserved, improved current, the extent to which $Z_V$ depends on the mass, momentum and Lorentz index is under investigation [9]. The analysis of the $D \to K^*$ matrix elements is also in progress.

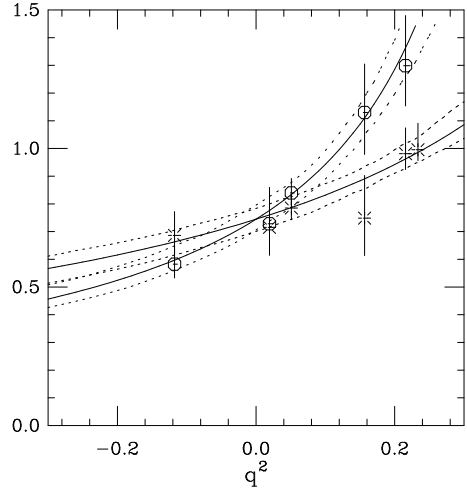

Figure 2. $f_K^+(q^2)/Z_V$ (circles), and $f_K^0(q^2)/Z_V$ (bursts), together with meson-dominance fits.

### Acknowledgements

This work was supported by SERC grant GR/H01069, and performed on a Meiko i860 Computing Surface supported by SERC grant GR/G32779, Meiko Limited and the University of Edinburgh. I thank Argonne National Laboratory for their hospitality and support over the summer.